\documentclass{aa}  
\usepackage[dvipsnames]{xcolor}
\usepackage{graphicx}
\usepackage{multirow}
\usepackage{booktabs}
\usepackage{array}
\usepackage{threeparttablex}
\usepackage{txfonts}
\usepackage{hyperref}

\definecolor{myblue}{RGB}{30,144,255}
\newcommand{\tasc}{$T_{\rm{asc}}$}
\newcommand{\porb}{$P_{\rm{orb}}$ }

\newcommand{\tzero}{$t_0$ }

\begin{document}

   \title{Evolution of the transitional millisecond pulsar \\ PSR J1023+0038 from Aqueye+ and NICER observations}
    \titlerunning{Evolution of PSR J1023+0038 from Aqueye+ and NICER observations} 

   \author{S. Conforti \fnmsep\thanks{\email{silvia.conforti@phd.unipd.it}}
          \inst{1,2,3}
          \and
          L. Zampieri \inst{2}
          \and
          M. Fiori \inst{2}
          \and
          A. Spolon \inst{2}
          \and 
          G. Naletto \inst{2,3} 
          \and 
          A. Burtovoi \inst{4,5}
          }
   
   \institute{Department of Astronomy, University of Padova, Via Vicolo dell'Osservatorio 3, 35122, Padova, Italy
         \and
             INAF - Osservatorio Astronomico di Padova, Via Vicolo dell'Osservatorio 5, 35122, Padova, Italy
        \and 
            Department of Physics and Astronomy, University of Padova, Via F. Marzolo 8, I-35131 Padova, Italy 
        \and 
            Università di Firenze - Dip. Fisica e Astronomia, Firenze, Italy
        \and 
            INAF - Osservatorio Astrofisico di Torino, Pino Torinese, Torino, Italy}

   \date{Received ...; accepted ...}

  \abstract 
   {Transitional millisecond pulsars are old, rapidly rotating neutron stars spun up by accretion from a low-mass companion star. These objects can switch between two emission regimes: rotation-powered radio pulsars and accreting X-ray pulsars. The mechanism responsible for these pulsations remains debated, with one prominent model suggesting that optical and X-ray pulsations arise from synchrotron emission produced by a shock resulting from the interaction between the pulsar wind and the accretion disk. This scenario is supported by the minimal phase lag between optical and X-ray pulses in PSR J1023+0038, and by their detection in the high-mode(flaring state) of the system.}
   {We present a new measurement of the phase lag between optical and X-ray pulse profiles of PSR J1023+0038 and investigate the recent evolution of the binary system (until 2023), monitoring the variation in the time of passage at the ascending node ($T_{asc}$).}
   {We performed a timing analysis of the optical observations of PSR J1023+0038 taken with Aqueye+ from 2021 through 2023 and  NICER X-ray observations taken in 2023. We used the optical and X-ray data to derive new measurements for the variation in $T_{asc}$ and employed simultaneous Aqueye+ and NICER observations to measure the phase lag between optical and X-ray pulses.}
   {We find the shift in $T_{asc}$ increases by $\sim$20 s each year, and in January 2023 we measured a phase lag between optical and X-ray data of $0.067 \pm 0.018$, or $112.3 \pm 30.7 \, \mu$s.}
   {After 2017, the variation in the time of passage at the ascending node follows a well-defined parabolic increasing trend, suggesting a corresponding increment in the orbital period and orbital separation of the system. This evolution is consistent with a scenario in which the binary system is evolving through nonconservative Roche lobe overflow and the donor is interacting with the pulsar wind, losing mass at a rate much higher than the inferred accretion rate. The measurement of the phase lag between optical and X-ray data confirms the common origin of such pulsations.}


   \maketitle

\section{Introduction}
Millisecond pulsars (MSPs) are old and fast-spinning neutron stars, spun up to periods shorter than 30 ms by the accretion of matter and angular momentum from a companion star \citep[][]{alpar1982new,radhakrishnan1982origin, salvo2021accretion}. Most are in low-mass binary systems ($< 1 M_{\odot}$), have weak magnetic fields ($\sim 10^8-10^9\, \rm{G}$), and are observed as accreting X-ray pulsars or as rotation-powered radio pulsars. Recent works identified MSPs that can switch between these two regimes, called transitional millisecond pulsars (tMPSs). In particular, these sources experience a transition of state on timescales of a few weeks, passing from a rotation-powered regime, where they act like radio pulsars with winds that prevent accretion onto their surface, to a regime where they accrete matter lost by the companion star, emitting a high amount of X-ray luminosity like X-ray binary systems do \citep[][]{papitto2021transitional}. Only three tMPSs are known so far, PSR J1023+0038 \citep[][]{archibald2009radio}, PSR J1227-4853 \citep[][]{de2010intriguing}, and PSR J1824-2452 \citep[][]{papitto2013swings}. 

PSR J1023+0038 was initially classified as a cataclysmic variable \citep[][]{bond2002first}, but between 2000 and 2001 observations showed the existence of an accretion disk, which soon disappeared, leaving a radio pulsar \citep{Thorstensen2005, archibald2009radio}. In 2013, the source underwent a new transition in its emission regime, with the disappearance of the radio pulsations, the reappearance of the accretion disk, and the onset of a strong double-peaked $\rm{H_{\alpha}}$ emission observed in the optical spectrum \citep[][]{halpern2013optical, patruno2013new, stappers2014state}. PSR J1023+0038 is also the only known tMPS that shows detectable UV \citep{miravalzanon2022, 2021jaodand} and optical pulsations; the latter were first discovered with SiFAP2 at the Telescopio Nazionale \textit{Galileo} \citep[][]{ambrosino2017optical} and then confirmed with Aqueye+ at the Copernicus telescope in Asiago \citep[][]{zampieri2019precise} and with the panoramic photometer-polarimeter mounted at BTA in Nizhniy Arkhyz \citep[][]{karpov2019study}. The emission mechanism of these optical pulsations is still under debate. Several models have been proposed to explain it \citep[][and reference therein]{campana2019probing, papitto2019pulsating, veledina2019pulsar}. In particular, the shock-driven mini-pulsar nebula scenario \citep[][]{veledina2019pulsar,papitto2019pulsating} postulates that optical and X-ray pulsations are produced by synchrotron emission caused by shocks that form when the pulsar wind meets the accretion disk. The fact that optical and X-ray pulses show a very short phase lag, between $0$ and $0.15$, and the fact that both appear only when the source is in the high-mode state, with X-ray pulsations in the high-mode first discovered by \cite{2015Archibald}, support this model and a common emission mechanism for the optical and X-ray pulsations \citep[][]{illiano2023investigating}.

We studied the evolution of the binary system, monitoring the variation in the time of passage at the ascending node, \tasc, and providing a more recent measurement of the phase lag between optical and X-ray pulse profiles. To achieve this, we performed a timing analysis of four Aqueye+ observing runs from 2021 through 2023, as well as of NICER X-ray data from 2023. We derived the timing solution of the source for the 2021 dataset, and we analyzed the evolution of \tasc, incorporating results from previous works \citep[][]{zampieri2019precise, burtovoi2020spin, illiano2023investigating, papitto2019pulsating, ambrosino2017optical, jaodand2016timing, archibald2009radio}. Finally, we considered a model for the evolution of the Roche lobe radius of the donor as a function of the mass loss rate, to account for the observed long-term behavior of \tasc.

In Sect. \ref{sec:datared} we present the observations and data reduction for the Aqueye+ and NICER observations. In Sect. \ref{sec:results} we describe the correction for the orbital motion and the results of our timing analysis. We discuss our results in Sect. \ref{sec.discussions}, and a summary follows in Sect. \ref{sec:conclusions}.

\section{Observations and data reduction}\label{sec:datared}
We carried out five observing runs of PSR J1023+0038 with the Aqueye+\footnote{\url{https://web.oapd.inaf.it/zampieri/aqueye-iqueye/index.html}} fast photon counter \citep[][]{barbieri2009aqueye, naletto2013aqueye, zampieri2015aqueye+} mounted at the 1.8 m Copernicus telescope (Asiago, Italy) in three different years, January 2021, January-February 2022, and January and December 2023 (see Table \ref{tab:oss_2021_info}).
The data were reduced with the \texttt{QUEST} software \citep[v. 1.1.5;][]{zampieri2015aqueye+}. The arrival time of each photon was referred to the Solar System barycenter using the \texttt{TEMPO2} package in TDB units \citep[][]{edwards2006tempo2, hobbs2006tempo2}, using JPL DE405 ephemerides. The position of PSR J1023+0038 was taken from \cite{2012Deller}: $\alpha= 10\rm{h}\,23\rm{m}\,47.687198\rm{s}$, $\delta= +00^\circ \,38'\,40.84551''$ at epoch MJD 54995.
We modified the barycentered time series, correcting the photon arrival times for the pulsar orbital motion. The orbital parameters are taken from \cite{2013archibald}: the projected semimajor axis $a\sin(i)=0.343356$ s and the orbital period \porb= 0.1980963155 d. The reference value for the time of passage at the ascending node is MJD 57449.7258 \citep[][]{ambrosino2017optical}. Variations in the time of passage at the ascending node are observed in PSR J1023+0038, the origin of which is not understood but clearly indicates variations in the orbital parameters.

Simultaneous NICER X-ray observations were requested for the Aqueye+ runs of January and December 2023. The corresponding log of the observations is reported in Table \ref{tab:nicer_obs}. For these observations, we applied the standard data processing procedure\footnote{\url{https://heasarc.gsfc.nasa.gov/docs/nicer/analysis_threads/}}, using the script \texttt{nicerl2} -- part of \texttt{HEASoft}\footnote{\url{https://heasarc.gsfc.nasa.gov/docs/software/heasoft/}} (v.6.34) software – with version 20240206 of the calibration files. The selected energy range for the analysis is $0.5-12\, \rm{keV}$. We finally barycentered the data with the script \texttt{barycorr}. Because of the known light-leak issue affecting NICER from May 2023, the observations from December 2023 presented here are only those in which a signal could be detected.

\section{Results}\label{sec:results}
We present a timing analysis of the new Aqueye+ optical and NICER X-ray data for PSR J1023+0038, from 2021 through 2023, and analyze the evolution of the passage at the ascending node, $T_{\rm{asc}}$, using observations gathered from 2017 through 2023 and incorporating results from previous works. 

\subsection{Evolution of the \tasc}
\label{evol_tasc}
The time of passage at the ascending node shows significant variations throughout the epochs. In this work, we report the measure of the \tasc \ offset for each Aqueye+ run, from January 2021 to December 2023. 

For each epoch, we performed an epoch folding search and corrected the barycentered time series for the orbital motion, using the orbital period reported in \cite{jaodand2016timing}, and varying only the \tasc \ to find the value that gives the highest signal, i.e., the pulsed profile with the highest $\chi^2$. We used the \tasc \ measured by \cite{ambrosino2017optical} as the reference value. Then, for each night, we explored a range of 40 values for the \tasc\ offset (20 in each direction from the expected value), with a time step of 0.5 s. The starting expected value was obtained from the best \tasc\ offset of the previous run, adding approximately 20 s to it. The final offset was determined by fitting the results of the epoch folding search with a Gaussian. The uncertainty on the best $\Delta T_{\rm{asc}}$ was calculated as the square root of the sum of the diagonal terms of the covariance matrix of the fit. The final best values of the \tasc\ are reported in Table \ref{tab:tab_tasc_values}, along with the final rotational period of the pulsar determined as described in Sect. \ref{timing_sol}. The values of \tasc\ were calculated with a trial pulsar rotational period $P_{\rm{init}}= 0.0016879874449$ s. 
\begin{table}[!ht]
    \caption{Aqueye+ runs of PSR J1023+0038 carried out between January 2021 and December 2023.}
    \centering
    \small
    \resizebox{\columnwidth}{!}{%
    \begin{tabular}{l c c c c}
       \toprule
       Aqueye+ run & \tzero (MJD) & $\Delta T_{\rm{asc}}$ (s) & $T_{\rm{asc}}$ (MJD) & $P_{\rm{spin}}$ (s) \\
       \midrule
       Jan.  2021  & 59226.0 & $44.35 \pm 0.17$  & $59226.055461$ &  0.0016879874467 $\pm$ 4.4$\times 10^{-13}$ \\
       Jan.  2023  & 59965.0 & $81.05 \pm 1.23$  & $59964.756622$ &  0.0016879874472 $\pm$ 9.0$\times 10^{-13}$ \\
       Dec. 2023  & 60290.0 & $97.33 \pm 1.39$  & $60290.030772$  &  0.0016879874474 $\pm$ 1.1$\times 10^{-12}$ \\
       \bottomrule
    \end{tabular}
    }
    \tablefoot{Aqueye+ runs of PSR J1023+0038, carried out at the 1.82 m Copernicus telescope. The reference epoch ($T_{\rm{ref}}$) corresponds to the start time of each run. The $\Delta T_{\rm{asc}}$ are calculated for the value reported in \cite{ambrosino2017optical}. The spin periods of January and December 2023 are those extrapolated from the long-term timing solution of \cite{burtovoi2020spin}, at the time of our observations.}
    \label{tab:tab_tasc_values}
\end{table}

\begin{figure}[!ht]
    \centering
    \includegraphics[width=0.9\linewidth]{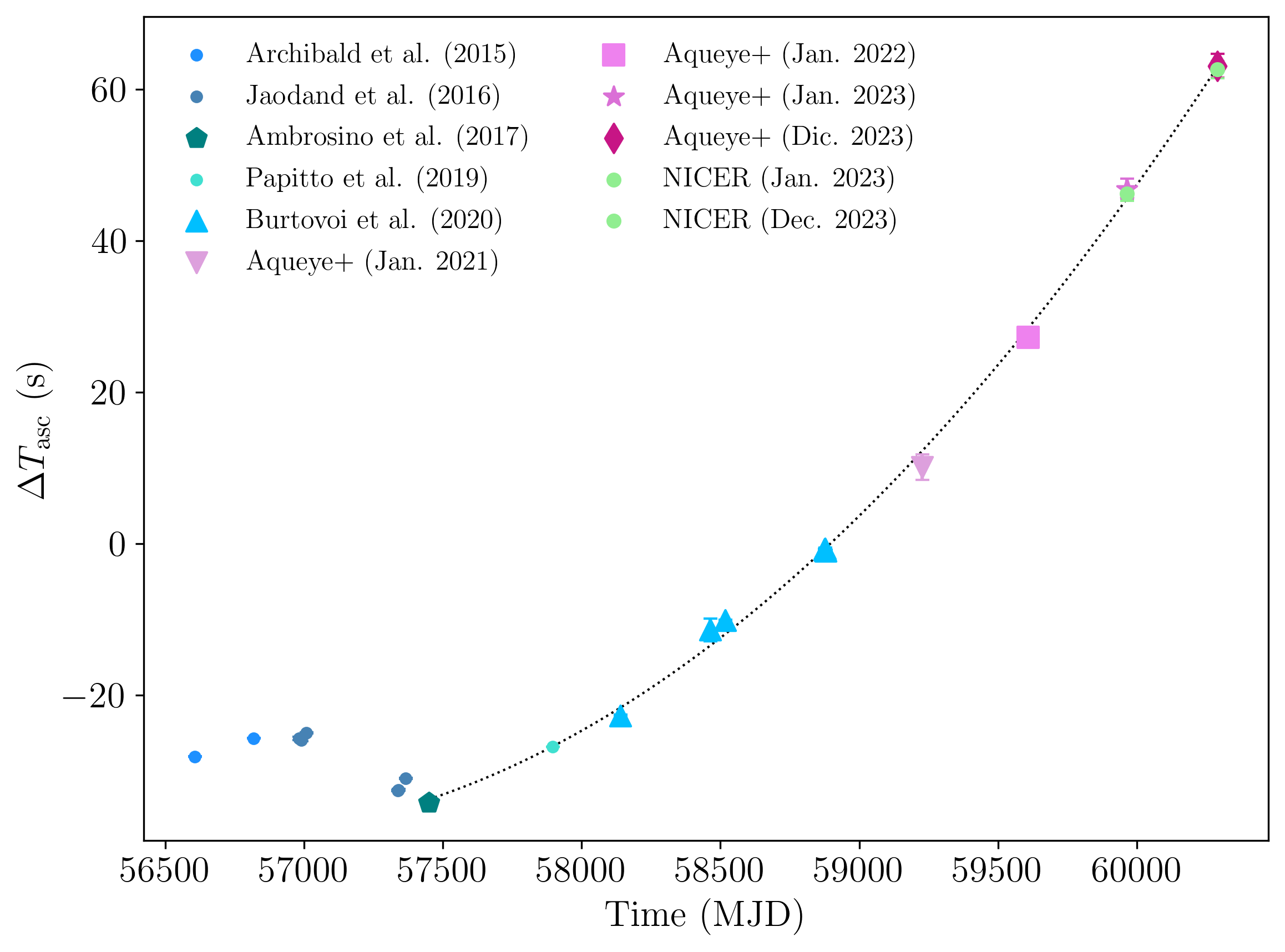}
    \caption{Variation in the time of passage at the ascending node, \tasc, concerning the reference value measured in radio reported in \cite{jaodand2016timing}. The dotted black line represents the fit of the data with Eq. \ref{fit_tasc_evolution}.}
    \label{fig:tasc_during_epochs}
\end{figure}

In Fig. \ref{fig:tasc_during_epochs} we show the \tasc\  evolution with time using our results and the \tasc\  offsets obtained from previous works \citep[][]{jaodand2016timing, ambrosino2017optical, papitto2019pulsating, burtovoi2020spin, illiano2023investigating}. We then performed a fit with a parabola \citep[][]{illiano2023investigating}:
\begin{equation}
    \Delta T_{\rm{asc}}({\rm{T}})=A + B\,{\rm{T}} + \frac{1}{2}C\,{\rm{T}^2},
\label{fit_tasc_evolution}
\end{equation}
where T is the time in MJD, and $A$, $B,$ and $C$ are the free parameters of the fit: $A=(2.46 \pm 0.17)\times10^{4}$ s, $B=-0.87 \pm 0.06\,\,\rm{s\, MJD^{-1}}$, and $C=(1.5 \pm 0.1)\times 10^{-5}\,\,\rm{s\, MJD^{-2}}$. The parameter $C$ can be expressed in s/s, $C'=(3.52\pm 0.23)\times10^{-11}$. It represents the derivative of the orbital period, and it is higher than the value found in \cite{illiano2023investigating} by a factor of $\sim 2$. 
We fitted only the optical data starting from 2017 \citep[][]{ambrosino2017optical}, when the variation in \tasc\  appears to be convincingly consistent with a parabolic trend.

\subsection{Folded pulse profiles for the single runs}
\label{timing_sol}
We report the timing solution for the Aqueye+ data of PSR J1023+0038 for the run of 2021, for which a sufficient number of nights is available to constrain it, and to this purpose, we corrected all the time series for the binary motion using the values of the \tasc \ (see Table \ref{tab:tab_tasc_values}) found from the previous analyses, and the orbital period and semimajor axis reported in \citet[see our Sect. \ref{sec:datared}]{jaodand2016timing}. We first folded the light curves of each night using a guess spin period $P_{\rm{init}}=0.0016879874449$ s and referring each folded light curve to a reference time \tzero $=59226.0$ MJD, i.e., midnight of January 20-21, 2021. The resulting pulsed profiles were fitted with the sum of two harmonically related sinusoids plus a constant \citep{ambrosino2017optical,zampieri2019precise,burtovoi2020spin}:
\begin{equation}
f(x) = K \{1 + A_1\sin(2\pi[x-x_1]) + A_2\sin(4\pi[x-x_2])\}, 
\label{two_harmonically_related_sinusoid}
\end{equation}
where $K$ is a normalization constant, $A_1$ is the coefficient of the first harmonic, $A_2$ is the coefficient of the second harmonic, and $x_1$ and $x_2$ are the phases of the first and second harmonics, respectively. The measured phases show a drift, $\Psi(t)$, with respect to a uniform rotational spin period, $P_{\rm{init}}$ \citep[see][]{2014MNRAS.439.2813Z}. To correct the spin period, we fitted this phase drift of the second harmonic with the following first-order polynomial:
\begin{equation}
\Psi(t) = \phi_0 + (\nu_0 - \nu_{init})(t-t_0), 
\label{first_order_polynomial}
\end{equation}
where $\phi_0$ and $\nu_0$ are the phase and the spin frequency at the reference time $t_0$, and $\nu_{\rm{init}}=1/P_{\rm{init}}$. 

\begin{figure}[!ht]
    \centering
    \includegraphics[width=0.9\linewidth]{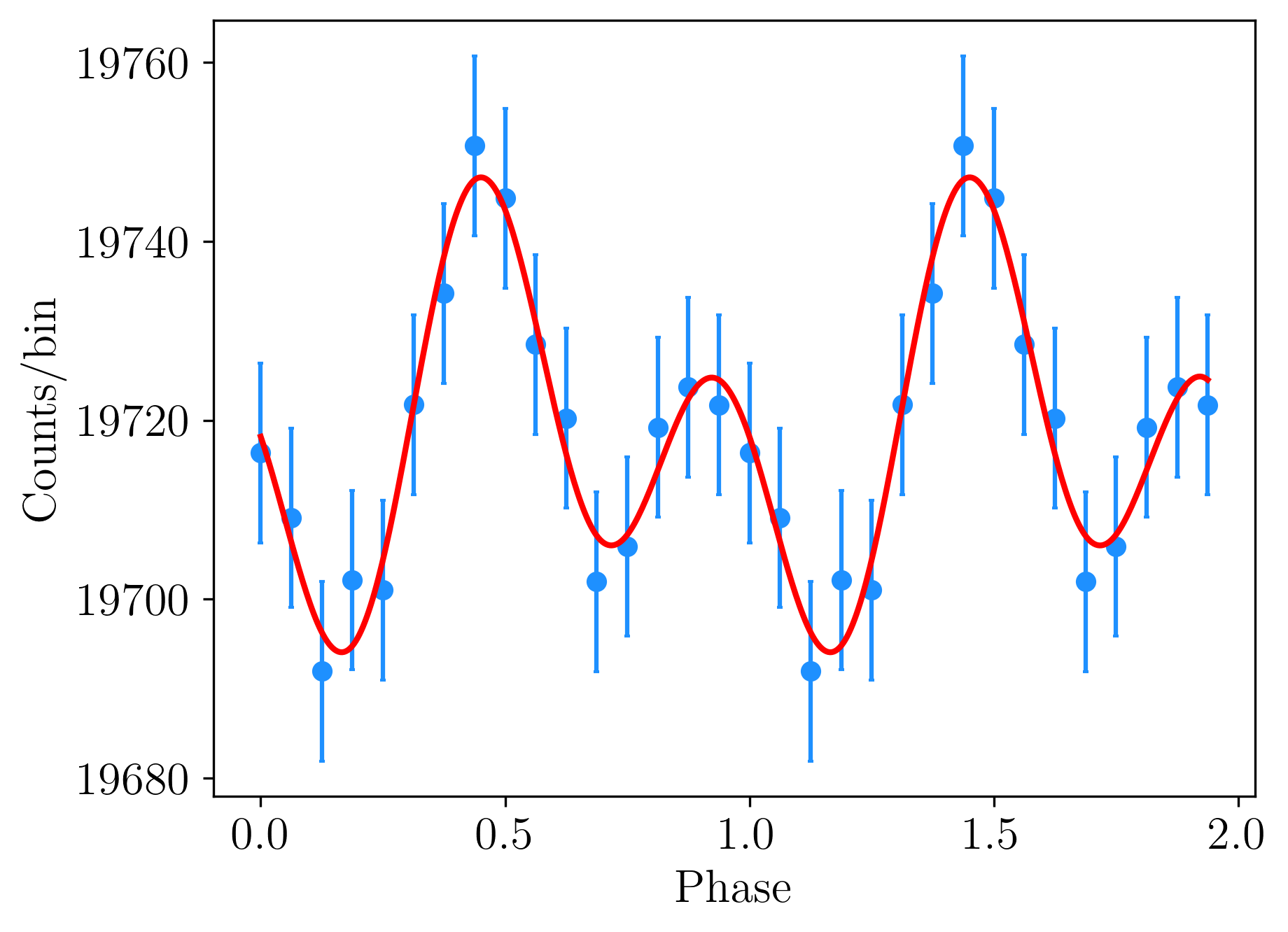}
    \caption{Optical pulse profile of PSR J1023+0038 obtained with Aqueye+ from January 12 to 16, 2021. The red line is the fit with Eq. \ref{two_harmonically_related_sinusoid}.}
    \label{fig:profile_final_2021}
\end{figure}

We report in Table \ref{tab:timing_sol_2021_e_2025} the timing solution for the January 2021 run. 
\begin{table}[!ht]
    \caption{Timing solution for PSR J1023+0038 from January 2021 Aqueye+ data.}
    \centering
    \begin{tabular}{l r }
    \hline
    \multicolumn{2}{c}{Aqueye+ Timing solutions} \\
    \hline
     & \multicolumn{1}{c}{January 2021}  \\ 
    \hline
    \vspace{0.005cm}\\
    $t_0$            &   59226  MJD                                      \\   
    $\phi_0$         &   0.799 $\pm$ 0.035                                \\  
    $\nu_0$          &   592.42146709 $\pm$ $1.7\times 10^{-7}$  Hz      \\  
    $P_{\rm{0}}$   &   0.0016879874473 $\pm$ $5\times10^{-13}$ s    \\  
    \hline
    \end{tabular}
    \label{tab:timing_sol_2021_e_2025}
\end{table}
We show the total pulsed profile of the source in Fig. \ref{fig:profile_final_2021}.

\begin{figure}[!ht]
    \centering
    \includegraphics[width=0.95\linewidth]{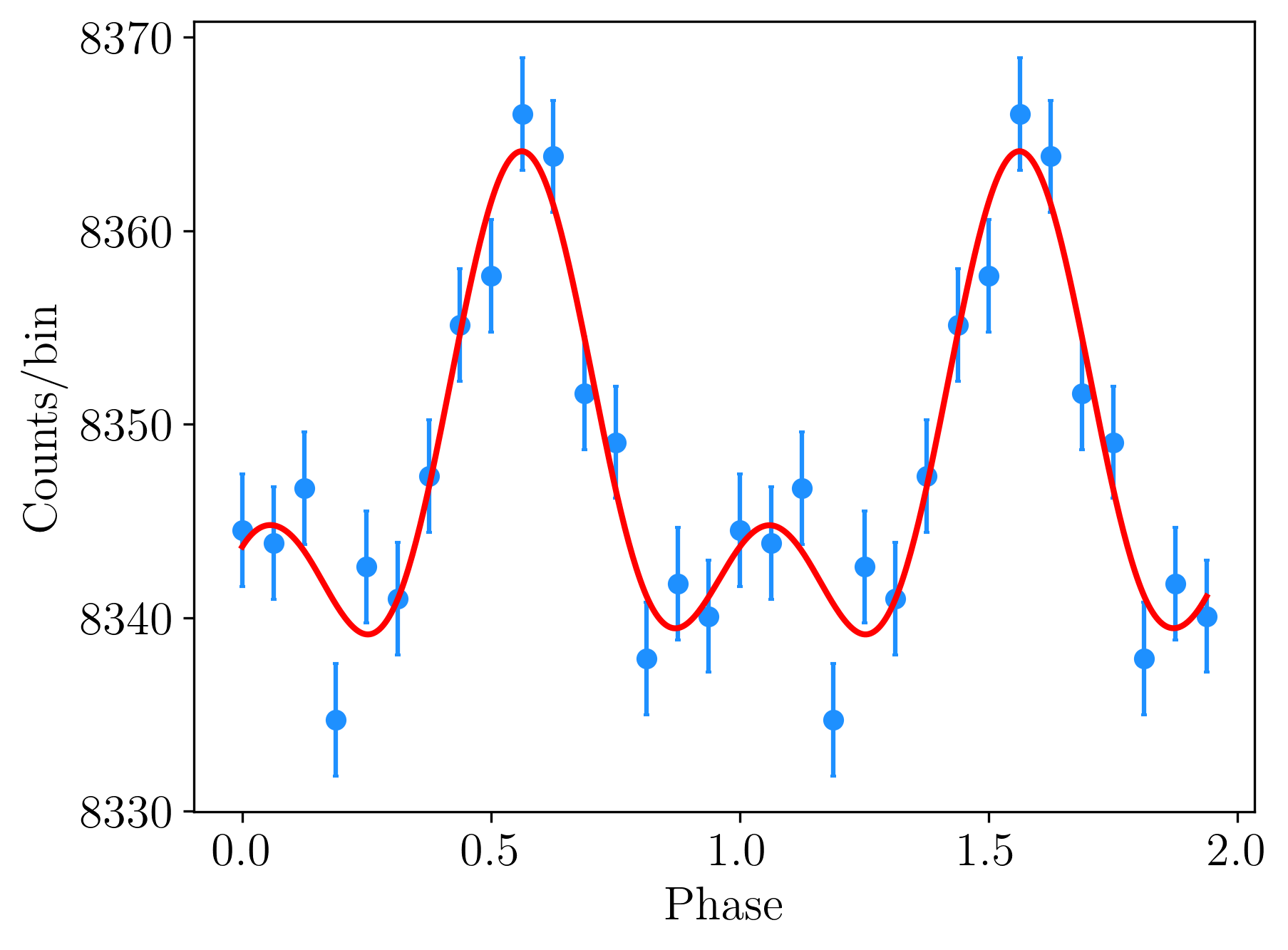}
    \caption{Optical pulse profile of PSR J1023+0038 obtained with Aqueye+ from December 12 and 15, 2023. The red line is the fit with Eq. \ref{two_harmonically_related_sinusoid}.}
    \label{fig:total_profile_dic_2023_ottico}
\end{figure}

The timing analysis for the 2022 Aqueye+ run is reported in \cite{illiano2023investigating}. For the January and December 2023 dataset, the number of nights is not sufficient to perform an analysis similar to that described above; hence, we corrected the data for the binary motion and folded them with the rotational period extrapolated from the accurate timing solution in \citet[values are reported in our Table \ref{tab:tab_tasc_values}]{burtovoi2020spin}, which differs from the spin period of 2018 by only $4.4\times 10^{-13}$ s. We performed the fit with Eq. \ref{two_harmonically_related_sinusoid} and show the final folded profile in Figs. \ref{fig:total_profile_dic_2023_ottico} and \ref{fig:shifts_dec_jan}.

\subsection{Time delay between optical and X-ray data}
We measured the delay between the time of arrival of the X-ray and optical pulses using the Aqueye+ and NICER data of January 2023 (Tables \ref{tab:oss_2021_info} and \ref{tab:nicer_obs}). We used the simultaneous NICER observation of January 21 to perform an epoch folding search to find the value of the \tasc\  offset (using only acquisitions when the source was in the high mode for $70-80\%$ of the time). We found an offset of $80.25 \pm 0.89$ s (computed taking the \tasc \ from \citealt{ambrosino2017optical} as the reference value). The offset is compatible within the uncertainties with that found from the optical data (Table \ref{tab:tab_tasc_values}). We corrected both the X-ray and optical datasets for binary motion  (considering only the optical observations within the X-ray high-mode window) with the \tasc\ offset of the X-ray data ($80.25$ s), the signals of which have a higher root mean square pulse amplitude. We also performed an epoch folding search using all the NICER observations reported in Table \ref{tab:nicer_obs} to find the \tasc\ offset for January and December 2023; we obtained $80.43 \pm 0.92\,\,\rm{s}$ and $96.86 \pm 1.13 \,\,\rm{s}$, respectively, which are consistent with the optical data (see Fig. \ref{fig:tasc_during_epochs}). 

\begin{figure}[!ht]
    \centering
    \includegraphics[width=0.8\linewidth]{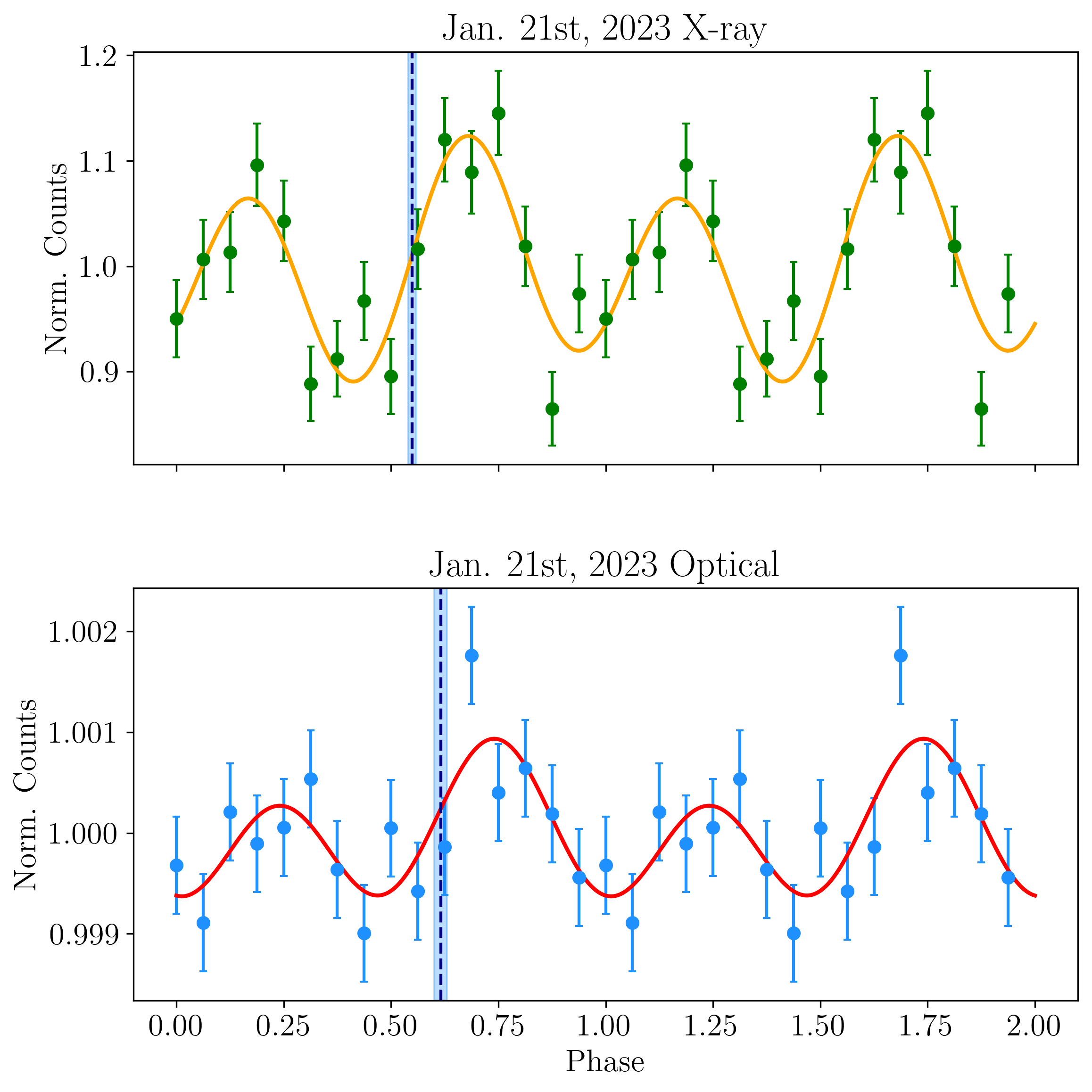}
    \caption{Pulse profiles of PSR J1023+0038 from January 21, 2023, in the optical (Aqueye+, bottom panel, blue dots) and X-ray (NICER, top panel, green dots) energy bands. Both profiles are fitted with the same model (yellow and red lines; see Eq. \ref{two_harmonically_related_sinusoid}), folded with the same spin period (see Table \ref{tab:tab_tasc_values}) and with the same \tzero=59965.0 MJD. The dashed navy lines in both panels represent the phase of the second harmonic ($x_2$), used to measure the phase lag, and the light blue shaded bands indicate the corresponding uncertainties.}
    \label{fig:shifts_dec_jan}
\end{figure}

Then, we folded the light curves of the simultaneous observations using the same \tzero (59965.0 MJD) and the spin period extrapolated from the \cite{burtovoi2020spin} long-term timing solution reported in Table \ref{tab:tab_tasc_values}. We then fitted both profiles with Eq. \ref{two_harmonically_related_sinusoid} and measured the related drift of the second harmonic, whose power spectral density is higher than that of the first harmonic. The optical pulses are found to be delayed with respect to the X-ray ones by $112.3 \pm 30.7\,\, \mu{\rm s}$, which corresponds to a phase lag of $0.067\pm0.018$. In Fig. \ref{fig:shifts_dec_jan} we show the X-ray and optical pulse profiles for January 21, 2023.

\section{Discussion}\label{sec.discussions}

\subsection{Evolution of \tasc}
In Sect. \ref{evol_tasc} we present the variation in the time of passage at the ascending node from 2021 through 2023. We report our results in Fig. \ref{fit_tasc_evolution}, alongside those obtained in previous works. We fitted the $\Delta T_{\rm{asc}}$ evolution with a second-order polynomial (see Eq. \ref{fit_tasc_evolution}), which is consistent with the observed trend of the data, in particular if we consider the $\Delta T_{\rm{asc}}$ values after the observations performed by \cite{ambrosino2017optical}. In the past, the variation in the \tasc\ was consistent with an irregular evolutionary scenario for the binary system; this scenario is further supported by its initial oscillating behavior after the onset of the accreting phase. Recent results, however, seem to support a different scenario, in which the increase in the \tasc could be driven by a rise in the orbital period and, thus, an increase in the separation between the companion star and the neutron star. In this section we discuss a possible general evolutionary scenario consistent with our recent results and the characteristics of this binary system.
 
Assuming accretion through Roche lobe overflow \citep{1993ARA&A..31...93V,2002apa..book.....F}, the exchange of mass leads to a variation in the distance between the two stars, due to the conservation of angular momentum, which translates into variations in the orbital period. Let us consider the general equation of the total angular momentum of the system, $J$, and its time derivative, $\dot{J}$, assuming a circular orbit: 
\begin{equation}
J=M_1M_2\sqrt{\frac{Ga}{M_1 + M_2}}M_{\odot}^{3/2},
\label{total_ang_mom}
\end{equation}
where $G$ is the gravitational constant, $a$ is the orbital separation, and $M_1$ and $M_2$ are the masses of the neutron star and the donor star, respectively. The time derivative of Eq. \ref{total_ang_mom} is
\begin{equation}
2 \frac{\dot{J}}{J} = \frac{\dot{a}}{a} +2 \left( \frac{\dot{M_1}}{M_1}+ \frac{\dot{M_2}}{M_2} \right) -\frac{\dot{M_1}+\dot{M_2}}{M_1+M_2} ,
\label{time_der_angmom}
\end{equation}
and the total mass and mass ratio of the system are defined as 
\begin{equation}
    M=M_1 + M_2 \,\,\,,\,\,\,\,\,q=\frac{M_2}{M_1}\,\,.
\label{total_mass}
\end{equation}
We used the equation that approximates the evolution of the Roche lobe radius of the donor star, $R_{2L}$, for $0.1\le q \le 0.8$, which is given by \citep[][]{frank2002accretion}
\begin{equation}
    \frac{\dot{R_{2L}}}{R_{2L}}=\frac{\dot{a}}{a} + \frac{\dot{M_2}}{3M_2} -\frac{1}{3}\frac{\dot{M}}{M},
\label{roche_lobe_radius2}
\end{equation}
The accretion luminosity in X-rays is $L_X=7\times10^{33}\ \rm{erg\ s^{-1}}$ in the high-mode state, and $L_X= 3\times10^{33}\ \rm{erg\ s^{-1}}$ in the low-mode state \citep[][]{2024A&A...690A.344M}. Neglecting emission from the neutron star surface, we estimated the mass accretion rate of the source as
\begin{equation}
L_{\rm{acc}} = \xi\frac{GM\dot{M}_{\rm{acc}}}{R_\star} \ \ \ \ ,
\label{lacc_real}
\end{equation}
where $G$ is the gravitational constant, $M$ the mass of the neutron star, $\dot{M}_{\rm{acc}}$ the mass accretion rate, $R_{\star}$ the radius of the neutron star, and $\xi$ the accretion efficiency, which depends on the inner radius, $R_{\rm{in}}$, of the accretion disk ($\xi=R_{\rm{\star}}/R_{in}$). In the case of PSR J1023+0038, in the high-mode state, material channeled along the magnetic field lines is likely not accreted onto the neutron star (as it would otherwise quench the pulsar wind) but rather expelled in a jet \citep[][]{2023A&A...677A..30B}. In the \cite{papitto2019pulsating} model, the inner radius of the disk, in the high-mode state, is beyond the light cylinder radius, at $\simeq80$; however,  when the source is in the low-mode state, the disk recedes up to a distance of $\simeq1600\,\rm{km}$, and most of the X-ray emission is produced by the compact jet along the pulsar rotational axis. Assuming $R\simeq80\,\rm{km}$, and considering the X-ray accretion luminosity in the high mode, we obtain
\begin{equation}
\dot{M}_{\rm{acc}}= 4.22 \times 10^{14}\left( \frac{M}{M_{\odot}}\right)^{-1} \ \rm{g \cdot s^{-1}}
.\end{equation}
For a conservative mass transfer, the value of the mass accretion rate is between $\dot{M_2}=-1.1\times 10^{19}\,\rm{g\,s^{-1}}$ (if $q=0.1$) and $\dot{M_2}=-2.5\times 10^{18}\,\rm{g\,s^{-1}}$ (if $q=0.8$; see \citealt{frank2002accretion}). These values, being at least four orders of magnitude higher than the value obtained from the observed X-ray luminosity, demonstrate the inconsistency of a conservative accretion scenario. 

The fact that the pulsar in PSR J1023+0038 is active implies that a powerful pulsar wind hits the donor surface, contributing to slowly evaporate the companion, as observed in other "redback binary systems." 
We can then consider a nonconservative case in which the accretion rate onto the neutron star is negligible and the mass loss is dominated by the donor, $\dot{M_1} \simeq \dot{M}_{\rm{acc}}<<-\dot{M}_2$, and $\dot{M}\simeq\dot{M_2}$. Then, Eq. \ref{time_der_angmom} assumes the following form:
\begin{equation}
2 \frac{\dot{J}}{J} = \frac{\dot{a}}{a} +2 \left(\frac{\dot{M_2}}{M_1} \right)  -\frac{\dot{M_2}}{M_1+M_2}\,\,. 
\label{tot_ang_mom_non_cons}
\end{equation}
Taking the time derivative of Kepler's third law, under the same assumptions, we obtain
\begin{equation}
\frac{\dot{a}}{a} = \frac{1}{3}\frac{\dot{M_2}}{M_2}\frac{M_2}{M} + \frac{2}{3}\frac{\dot{P}}{P} ,
\label{a_vs_porb_non_cons2}
\end{equation}
where $P$ is the orbital period. Since we do not know exactly how the companion star evolves as it interacts with the pulsar wind, we considered a general case in which the radius of the donor evolves as
\begin{equation}
\frac{\dot{R_{2}}}{R_{2}} = \alpha \frac{\dot{M_2}}{M_2}\,\,.
\label{very_slowly_r2_vs_m2}
\end{equation}
When $\alpha=1$, this relation represents the evolution of a low-mass donor that is losing mass in a quasi-steady state. Assuming that the system is in (or close to) Roche lobe overflow, and hence $R_2\simeq R_{\rm{2L}}$, we substituted this expression into Eq. \ref{roche_lobe_radius2} and then combined it with Eq. \ref{a_vs_porb_non_cons2}. The resulting relation is
\begin{equation}
\frac{\dot{P}}{P} = \frac{3}{2}\beta\frac{\dot{M_2}}{M_2} \,\,\,\,\,\,\,\,\,\,\,\, ,\,\,\,\,\,\, \rm{where}\,\,\,\,\,\,\,\,\, \beta=\left(\alpha - \frac{1}{3} \right)\,\,\, .
\label{third_case_non_cons_compata}
\end{equation}
As mentioned above, the case $\alpha=1$ represents the evolution of a low-mass donor in a quasi-steady state. This is not a plausible scenario since $\beta=-2/3$ and ${\dot{P}}/P={\dot{M}_2}/M_2 < 0$, which is at variance with the observed increase in the orbital period. The same conclusion holds in general for $\alpha > 1/3$, while for $\alpha < 1/3$, ${\dot{P}}/P \propto -{\dot{M}_2}/M_2 > 0$, which is consistent with the observations. We note that the case $\alpha=0$ ($\beta=1/3$) would represent the evolution of a system with a constant donor and Roche lobe radius. In these conditions, the mass-loss rate of the donor would be $-\dot{M}_2/M_2 = (1/2)(\dot{P}/P) = 8.2 \times 10^{18}\, \rm{gs^{-1}}$, higher by four orders of magnitude than the mass accretion rate. As we do not know the physics of the interaction between the pulsar wind and the donor, different values of $\alpha$ would be possible --- implying different evolutions for the donor and different values for the mass-loss rate --- but they must be on the same order of magnitude as the previous estimate for $|\alpha| \approx 1$. At this pace, the donor would be completely consumed in $\sim 8$ M yrs. Thus, a scenario with a binary system evolving through nonconservative Roche lobe overflow, similar to the one presented above, appears to be consistent with the observed steady parabolic increase in the orbital period with time.

\subsection{Time delay between optical and X-rays}
We analyzed NICER and Aqueye+ data from January 2023 to perform new measurements of the time shift between the optical and X-ray pulses of PSR J1023+0034.  
We managed to measure the shift simultaneously between optical and X-rays only for the January 2023 data (January 21) because for the December 2023 dataset, the duration of the simultaneous X-ray observation (December 15) was too short. The time lag between optical and X-ray pulses is $0.036\pm0.016$, which corresponds to $60.7\pm27.7\,\rm{\mu s}$. This result is consistent with that of \cite{illiano2023investigating}, who report that, from the second harmonic of the pulsed profiles, the time lag between optical and X-ray pulses lies in the range $(0-250)\,\rm{\mu s}$. This, along with the fact that the optical and X-ray pulses have similar morphologies, confirms that they are produced in the same region and with the same emission mechanism, as proposed by \cite{papitto2019pulsating} and \citet[see the references therein]{veledina2019pulsar}. They present a possible scenario to explain such a simultaneous emission, the so-called striped wind model. In this scenario, the source is in the disk state but still able to emit a relativistic wind. In a region a few light cylinder radii from the star, the pulsar wind interacts with the accretion disk, producing two shocked symmetric regions; this is where synchrotron radiation is produced and pulsed emission in the X-ray and optical bands originates. They also estimated the synchrotron timescale for optical photons, obtaining a value compatible with the lag observed between optical and X-ray photons.

\section{Summary}\label{sec:conclusions}
We have studied the evolution of the PSR J1023+0038 binary system, monitoring the long-term variation in the time of passage at the ascending node, \tasc, and providing new measurements of the phase lag between the optical and X-ray pulse profiles. We performed a timing analysis of three runs of Aqueye+ optical data from 2021 through 2023, as well as NICER X-ray data from 2023. We derived the timing solution of the source for January 2021, and we analyzed the evolution of the \tasc, including also results from previous works.
We measured the shift between optical and X-ray pulses using simultaneous Aqueye+ optical data and NICER X-ray data from January 2023. 
The phase lag measured is $0.036 \pm0.016$, in agreement with \cite{illiano2023investigating} and extending the baseline of these measurements by one year. This result further confirms the common origin and emission mechanism of X-ray and optical pulses. Concerning the evolution of the \tasc, starting from 2017, it follows a parabolic increasing trend, which translates into an increase in the orbital period and orbital separation of the system. 
A scenario of a binary system evolving through nonconservative Roche lobe overflow in which the donor is interacting with the pulsar wind, losing mass at a rate much higher than the inferred accretion rate, appears to be consistent with the observed steady increase in the orbital period of the system. The large majority of the mass stripped from the donor was likely shocked by the pulsar wind itself and ejected from the system after crossing the Roche lobe. Further optical and X-ray timing observations of PSR J1023+0038 are needed to continue to monitor the evolution of the parameters of the binary system and to determine the long-term rotational history of the pulsar.

\begin{acknowledgements}
      SC is supported by the INAF Doctorate program. SC and LZ acknowledge financial support from the INAF Research Grant “Uncovering the optical beat of the fastest magnetized neutron stars (FANS)”. Based on observations collected at the Copernico 1.82m telescope (Asiago Mount Ekar, Italy), INAF - Osservatorio Astronomico di Padova. This work has made use of data products and software provided by the High Energy Astrophysics Science Archive Research Center (HEASARC), a service of the Astrophysics Science Division at NASA/GSFC and the High Energy Astrophysics Division of the Smithsonian Astrophysical Observatory. We would like to express our thanks to Alessandro Papitto, Giulia Illiano, Arianna Miraval Zanon for the very interesting discussions on the content of this article and for his support in training S.C. on pulsar timing techniques.
\end{acknowledgements}

\bibliographystyle{aa}
\bibliography{bibliography}

\appendix
\section{Observations}

\begin{table}[!ht]
    \caption{Aqueye+ observations of PSR J1023+0038 from January 2021 to December 2023.}
    \centering
    \scriptsize
    \resizebox{\columnwidth}{!}{ 
    \begin{tabular}{c >{\raggedleft\arraybackslash}p{3cm} c r}
    \toprule
    Observation ID   & \multicolumn{1}{r}{Start Time (UTC)} & Start Time (MJD) & Duration (s) \\
    \midrule
    \multicolumn{4}{c}{\footnotesize{Aqueye+ Optical Data}} \\
    \midrule
    QEYE$\_$20210112-022231  & Jan. 12, 2021 01:29:24.778 & 59226.062092 & 1798.663 \\
    QEYE$\_$20210112-030013  & 02:07:05.713 & 59226.088260 & 1798.633 \\
    QEYE$\_$20210112-033504  & 02:41:56.956 & 59226.112464 & 1799.053 \\
    QEYE$\_$20210112-041003  & 03:16:57.119 & 59226.136772 & 1799.441 \\
    QEYE$\_$20210112-044613  & 03:53:07.535 & 59226.161892 & 1799.062 \\
    QEYE$\_$20210112-052324  & 04:30:18.917 & 59226.187718 & 1799.048 \\
    QEYE$\_$20210112-055805  & 05:05:00.102 & 59226.211806 & 898.946  \\
    QEYE$\_$20210113-023449  & Jan. 13, 2021 01:41:48.080 & 59227.070695 & 1798.449 \\
    QEYE$\_$20210113-030825  & 02:15:24.161 & 59227.094029 & 1798.656 \\
    QEYE$\_$20210113-034203  & 02:49:02.057 & 59227.117384 & 1799.250 \\
    QEYE$\_$20210113-041733  & 03:24:32.436 & 59227.142042 & 1799.179 \\
    QEYE$\_$20210113-045130  & 03:58:29.992 & 59227.165624 & 1799.019 \\
    QEYE$\_$20210113-052515  & 04:32:15.278 & 59227.189065 & 1798.947 \\
    QEYE$\_$20210113-055900  & 05:06:00.408 & 59227.212504 & 898.935  \\
    QEYE$\_$20210114-023019  & Jan. 14, 2021 01:37:24.769 & 59228.067647 & 1798,919 \\
    QEYE$\_$20210114-030423  & 02:11:28.018 & 59228.091296 & 1798,919 \\
    QEYE$\_$20210114-034050  & 02:47:55.149 & 59228.116610 & 1799,128 \\
    QEYE$\_$20210114-041437  & 03:21:42.531 & 59228.140075 & 1799,448 \\
    QEYE$\_$20210114-044832  & 03:55:37.890 & 59228.163632 & 1799,323 \\
    QEYE$\_$20210114-052236  & 04:29:42.140 & 59228.187293 & 1798,765 \\
    QEYE$\_$20210114-055632  & 05:03:38.144 & 59228.210858 & 898,920 \\
    QEYE$\_$20210116-031012  & Jan. 16, 2021 02:17:29.737 & 59230.095483 & 1799.314 \\
    QEYE$\_$20210116-034531  & 02:52:48.148 & 59230.120001 & 1799.044 \\
    QEYE$\_$20210116-042150  & 03:29:08.508 & 59230.145237 & 1798.887 \\
    QEYE$\_$20210116-045712  & 04:04:30.720 & 59230.169799 & 1799.143 \\
    QEYE$\_$20210116-053219  & 04:39:36.737 & 59230.194175 & 1798.975 \\
    QEYE$\_$20220129-014645  & Jan. 29, 2022 00:55:06.246 & 59608.038267 & 1799.43 \\
    QEYE$\_$20220129-021723  & 01:25:43.430 & 59608.059530 & 1799.43 \\
    QEYE$\_$20220129-024810  & 01:56:30.724 & 59608.080911 & 1799.43 \\
    QEYE$\_$20220129-032714  & 02:35:35.497 & 59608.108050 & 1799.05 \\
    QEYE$\_$20220129-035755  & 03:06:17.308 & 59608.129367 & 1799.43 \\
    QEYE$\_$20220129-042915  & 03:37:37.355 & 59608.151127 & 1799.43 \\
    QEYE$\_$20220129-050002  & 04:08:23.269 & 59608.172491 & 1799.43 \\
    QEYE$\_$20220130-010757  & Jan. 30, 2022 00:16:21.299 & 59609.011358 & 1799.43 \\
    QEYE$\_$20220130-014125  & 00:49:50.371 & 59609.034611 & 1799.43 \\
    QEYE$\_$20220130-021621  & 01:24:45.800 & 59609.058863 & 1799.26 \\
    QEYE$\_$20220130-024724  & 01:55:49.132 & 59609.080430 & 1799.24 \\
    QEYE$\_$20220130-032433  & 02:32:58.415 & 59609.106232 & 1799.3 \\
    QEYE$\_$20220130-035525  & 03:03:51.420 & 59609.127678 & 1799.43 \\
    QEYE$\_$20220130-042708  & 03:35:33.409 & 59609.149692 & 1799.43 \\
    QEYE$\_$20220130-231911  & 22:27:39.820 & 59609.935877 & 1799.21 \\
    QEYE$\_$20220130-235112  & 22:59:40.488 & 59609.958108 & 1799.43 \\
    QEYE$\_$20220131-005611  & Jan. 31, 2022 00:04:40.260 & 59610.003244 & 1799.43 \\
    QEYE$\_$20220131-013048  & 00:39:16.346 & 59610.027273 & 1799.43 \\
    QEYE$\_$20220131-020246  & 01:11:14.682 & 59610.049475 & 1799.34 \\
    QEYE$\_$20220131-023337  & 01:42:05.895 & 59610.070902 & 1799.43 \\
    QEYE$\_$20220202-010515  & Feb. 2, 2022 00:13:50.934 & 59612.009617 & 1799.37 \\
    QEYE$\_$20220202-014053  & 00:49:29.535 & 59612.034370 & 1799.05 \\
    QEYE$\_$20220203-005757  & Feb. 3, 2022 00:06:37.468 & 59613.004600 & 1799.43 \\
    QEYE$\_$20220203-012943  & 00:38:23.091 & 59613.026656 & 1799.07 \\
    QEYE$\_$20220203-020723  & 01:16:04.096 & 59613.052825 & 1799.43 \\
    QEYE$\_$20220203-023800  & 01:46:40.636 & 59613.074081 & 1799.12 \\
    QEYE$\_$20220203-031435  & 02:23:15.591 & 59613.099486 & 1799.26 \\
    QEYE$\_$20220203-034549  & 02:54:29.691 & 59613.121177 & 1799.07 \\
    QEYE$\_$20230121-012433  & Jan. 21, 2023 00:32:14.118 & 59965.022386 & 1799.1 \\
    QEYE$\_$20230121-015546  & 01:03:27.477 & 59965.044068 & 1799.04 \\
    QEYE$\_$20230121-023109  & 01:38:50.496 & 59965.068640 & 1799.41 \\
    QEYE$\_$20230121-033656  & 02:44:38.231 & 59965.114331 & 1799.09 \\
    QEYE$\_$20231212-014334  & Dec. 12, 2023 00:46:27.276 & 60290.032260 & 1798.773 \\
    QEYE$\_$20231212-021736  & 01:20:29.214 & 60290.055893 & 3599.222 \\
    QEYE$\_$20231212-032349  & 02:26:43.528 & 60290.101893 & 1799.328 \\
    QEYE$\_$20231212-035751  & 03:00:45.943 & 60290.125532 & 3599.406 \\
    QEYE$\_$20231212-050313  & 04:06:08.712 & 60290.170934 & 1798.973 \\
    QEYE$\_$20231212-053409  & 04:37:04.884 & 60290.192417 & 1798.899 \\
    QEYE$\_$20231215-015007  & Dec. 15, 2023 00:53:26.054 & 60293.037107 & 1799.301 \\
    QEYE$\_$20231215-022539  & 01:28:57.344 & 60293.061775 & 607.544 \\
    QEYE$\_$20231215-031850  & 02:22:09.147 & 60293.098717 & 1798.976 \\
    QEYE$\_$20231215-035152  & 02:55:12.259 & 60293.121696 & 1799.171 \\
    QEYE$\_$20231215-042249  & 03:26:08.911 & 60293.143159 & 1798.632 \\
    QEYE$\_$20231215-045716  & 04:00:36.745 & 60293.167092 & 1799.169 \\
    \bottomrule
    \end{tabular}}
    \label{tab:oss_2021_info}
\end{table}

\begin{table*}[!ht]
     \caption{NICER X-ray observations of PSR J1023+0038 from January 20 to December 30, 2023.}
    \centering
    \scriptsize
    \begin{tabular}{c >{\raggedleft\arraybackslash}p{3cm} c c r }
    \toprule
    Observation ID   & \multicolumn{1}{r}{Start Time (UTC)} & Start Time (MJD) & End Time (MJD) & Duration \tablefootmark{a} (s)\\
    \midrule
    \multicolumn{5}{c}{\footnotesize{NICER X-ray Data}} \\
    \midrule
    5034060101 & Jan. 20, 2023 19:23:29 & 59964.808752 & 59964.948583 & 2846  \\
    5034060102 & Jan. 21, 2023 00:01:53 & 59965.001308 & 59965.981259 & 4221  \\
    5034060103 & Jan. 22, 2023 00:48:20 & 59966.033565 & 59966.964547 & 8649  \\
    5034060104 & Jan. 23, 2023 21:50:22 & 59967.002062 & 59967.932244 & 8964  \\
    5034060105 & Jan. 24, 2023 19:35:20 & 59968.042594 & 59968.835403 & 10077 \\
    5034060106 & Jan. 25, 2023 21:42:19 & 59969.010523 & 59969.918461 & 14733 \\
    5034060107 & Jan. 26, 2023 00:48:06 & 59970.034180 & 59970.174872 & 3830  \\
    6034060104 & Dec. 14, 2023 21:55:49 & 60292.913764 & 60292.916848 & 265   \\
    6034060105 & Dec. 15, 2023 02:36:24 & 60293.108613 & 60293.110349 & 149   \\
    6034060106 & Dec. 16, 2023 03:21:36 & 60294.140002 & 60294.141540 & 132   \\
    6034060107 & Dec. 17, 2023 22:39:34 & 60295.944145 & 60295.948157 & 345   \\
    6034060108 & Dec. 18, 2023 00:12:55.053 & 60296.008984 & 60296.989498 & 3439 \\
    6034060109 & Dec. 19, 2023 00:58:24.468 & 60297.040592 & 60297.957787 & 3280 \\
    6034060110 & Dec. 20, 2023 00:13:00.322 & 60298.009053 & 60298.990252 & 5636 \\
    6034060111 & Dec. 21, 2023 00:59:08.636 & 60299.041079 & 60299.958039 & 2561 \\
    6034060112 & Dec. 22, 2023 00:14:41.391 & 60300.010210 & 60300.990190 & 6354 \\
    6034060113 & Dec. 23, 2023 04:05:16.602 & 60301.170349 & 60301.952179 & 2152 \\
    6034060114 & Dec. 24, 2023 00:15:41.227 & 60302.010905 & 60302.991047 & 2434 \\
    6034060115 & Dec. 25, 2023 00:59:56.287 & 60303.041634 & 60303.959167 & 3667 \\
    6034060116 & Dec. 26, 2023 00:13:45.797 & 60304.009574 & 60304.217321 & 4892 \\
    6034060117 & Dec. 27, 2023 22:39:24.615 & 60305.944065 & 60305.959518 & 1334 \\
    6034060118 & Dec. 28, 2023 00:12:15.104 & 60306.008567 & 60306.991997 & 2098 \\
    6034060120 & Dec. 30, 2023 00:33:11.253 & 60308.023057 & 60308.992345 & 3242 \\
    \bottomrule
    \end{tabular}
    \label{tab:nicer_obs}
    \tablefoot{\tablefoottext{a}{Duration of the effective observing time after applying the standard filtering. Owing to the light-leak problem affecting NICER in December 2023, the time that was discarded on average is quite significant ($\sim20\%$).}
    }
\end{table*}

\end{document}